\documentclass[fleqn,usenatbib]{mnras}

\usepackage[T1]{fontenc}

\DeclareRobustCommand{\VAN}[3]{#2}
\let\VANthebibliography\thebibliography
\def\thebibliography{\DeclareRobustCommand{\VAN}[3]{##3}\VANthebibliography}

\usepackage{graphicx}	
\usepackage{amsmath}	
\usepackage{amssymb}	
\usepackage{multirow}
\usepackage{ulem}

\usepackage{newtxtext,newtxmath}

\title[A persistent double nuclear structure in 3C\,84]{A persistent double nuclear structure in 3C\,84}

\author[J. Oh et al.]{Junghwan Oh$^{1,2,3}$,\thanks{E-mail: joh@sejong.ac.kr, jhodgson@sejong.ac.kr}
Jeffrey A. Hodgson$^{1,2}$,
Sascha Trippe$^{3,4}$,
Thomas P. Krichbaum$^{5}$,
Minchul Kam$^{3}$,\newauthor
Georgios Filippos Paraschos$^{5}$,
Jae-Young Kim$^{2,5}$,
Bindu Rani$^{2,6}$,\newauthor
Bong Won Sohn$^{2}$,
Sang-Sung Lee$^{2}$,
Rocco Lico$^{7,5}$,
Elisabetta Liuzzo$^{8}$,
Michael Bremer$^{9}$,
Anton Zensus$^{5}$,
\\
$^{1}$Department of Physics and Astronomy, Sejong University, 209 Neungdong-ro, Gwangjin-gu, Seoul 05006, Korea\\
$^{2}$Korea Astronomy and Space science Institute, 776 Daedeokdae-ro, Yuseong-gu, Daejeon, 30455, Korea \\
$^{3}$Department of Physics and Astronomy, Seoul National University, 1 Gwanak-ro, Gwanak-gu, Seoul 08826, Korea \\
$^{4}$SNU Astronomy Research Center, Seoul National University, Gwanak-gu, Seoul 08826, Korea\\
$^{5}$Max-Planck-Institut f\"{u}r Radioastronomie, Auf dem H\"{u}gel 69, 53121, Bonn, Germany\\
$^{6}$Department of Physics, American University, Washington, DC 20016, USA \\
$^{7}$Instituto de Astrof\'{\i}sica de Andaluc\'{\i}a, IAA-CSIC, Apdo. 3004, 18080 Granada, Spain.
 \\
$^{8}$ INAF-Istituto di Radioastronomia and Italian ALMA Regional Centre, Via P. Gobetti 101, I-40129 Bologna, Italy \\
$^{9}$Institut de radioastronomie millim\'{e}trique, 300 rue de la Piscine, Domaine Universitaire 38406 Saint Martin d'H\`{e}res, France \\
}

\date{Accepted 2021 October 18. Received 2021 October 15; in original form 2021 March 11}

\pubyear{2021}

\begin{document}
\label{firstpage}
\pagerange{\pageref{firstpage}--\pageref{lastpage}}
\maketitle

\begin{abstract}
3C\,84 (NGC\,1275) is the radio source at the center of the Perseus Cluster and exhibits a bright radio jet. We observed the source with the Global Millimeter VLBI Array (GMVA) between 2008 and 2015, with a typical angular resolution of $\sim$50\,$\mu$as. The observations revealed a consistent double nuclear structure separated by $\sim$770 gravitational radii assuming a Black Hole mass of 3.2 $\times 10^{8}$ $M_{\odot}$. 
The region is likely too broad and bright to be the true jet base anchored in the accretion disk or Black Hole ergosphere. 
A cone and parabola were fit to the stacked (time averaged) image of the nuclear region. 
The data did not strongly prefer either fit, but combined with a jet/counter-jet ratio analysis, an upper limit on the viewing angle to the inner jet region of $\leq$35$^{\circ}$ was found. This provides evidence for a variation of the viewing angle along the jet (and therefore a bent jet) within $\sim$0.5 parsec of the jet launching region.
In the case of a conical jet, the apex is located $\sim$2400 gravitational radii upstream of the bright nuclear region and up to $\sim$600 gravitational radii upstream in the parabolic case. We found a possible correlation between the brightness temperature and relative position angle of the double nuclear components, which may indicate rotation within the jet. 
\end{abstract}




\section{Introduction}
Jets from supermassive Black Holes (SMBHs), at the center of active galactic nuclei (AGN), have been extensively studied over the past few decades.  The jet launching mechanism(s) is still a matter of debate. In general, there are two theoretical models describing how  jets can be formed.  Magnetic fields (B-fields) play an important role in  jet launching, where twisted and wrapped B-field geometry guides the material outwards. In the Blandford-Payne mechanism \citep[][BP hereafter]{BP},  B-fields are anchored in the accretion disk and form a light-cylinder in the shape of a jet. In the Blanford-Znajek model \citep[][BZ hereafter]{BZ}, the jet is powered by the rotation energy of the Black Hole, and is launched from its ergosphere. Observationally,  the innermost jet morphology, such as jet width or the shape of the nozzle, could distinguish between the two models. High-resolution VLBI observations allow us to probe the central region of an AGN on scales down to a few tens of  gravitational radii. The observational differences between the jet launching models are  resolvable at these scales.

The Global mm-VLBI Array (GMVA) offers a unique ability to unravel the physical processes in the immediate vicinity of the central black hole of an AGN. 
We present here a detailed analysis of the GMVA images of a nearby \citep[$z=0.017559$, ][]{strauss1992} radio galaxy, 3C\,84. At the redshift 0.0176 the angular scale of 1 mas corresponds to a spatial scale of 0.35~parsec~mas$^{-1}$, where we adopt for the cosmological constants $H_{0}$ = 67.7 km/s/Mpc, $\Omega_{\Lambda}$ = 0.693 and $\Omega_{M}$ = 0.307 \citep{planck}. We adopt 3.2 $\times 10^{8}$ $M_{\odot}$ for the mass of the SMBH in 3C\,84 \citep[][]{park2017}. An angular resolution of $\sim$50\,$\mu$as corresponds to $\sim$550 gravitational radii($R_g$) in the source frame.

Being nearby and bright, 3C\,84 is one of the most extensively studied radio galaxies. Although it is further away and has a smaller SMBH, it is nevertheless often compared with M\,87. In both cases, we can directly observe close to the jet launching region, allowing us to compare and contrast the jet launching mechanisms. In VLBI images, 3C\,84 shows a south oriented core-jet structure, with the expanding jet being transversely resolved. Previous VLBI imaing at 43 GHz revealed 3 distinct jet features, named C1, C2 and C3 \citep[e.g.,][]{nagai2014}. The region, C1, is the compact core area in the north of the structure, that is usually the map center. A weak and diffuse emission region at around 2 mas apart from the core to the south-west is C2. C3 is a slowly moving bright jet component located at $\sim$2.5 mas south from the core and may be the source of $\gamma$-ray emission \citet{hodgson2018b, hodgson2021}. In this paper, we are investigating C1, the nuclear region. Early results of this study are presented in \citet{hodgson2018a}.

A recent paper by \citet{giovannini2018} used the orbiting VLBI experiment RadioAstron to observe 3C\,84 at comparable angular resolution to the GMVA. The authors found a very wide jet base which converged onto a central bright component. This broad base was $\sim$250\,$R_{S}$ wide. According to the authors, the separation between jet base and SMBH is $\sim$350\,$R_{S}$ (adopting $M_{\rm BH} = 2 \times 10^9 M_{\odot}$.)
While a parabolic expanding jet could not be ruled out, it has been suggested that the jet expands conically from a radius of $\sim$250\,$R_{S}$ and is launched from the disk. Note that, the black hole mass in 3C\,84 is uncertain with mass estimates ranging from $10^7$ \citep{onori2017} to $10^9$ \citep{scharwachter2013}. If we chose an average value of 3 $\times 10^8$ as published by \citet{park2017}, the RadioAstron data  suggest the jet base to be even wider, $\sim$1000\,$R_{S}$.

Previous studies \citep[e.g.,][]{krichbaum1992, dhawan1998, suzuki2012, hodgson2021} revealed that the sub-parsec scale kinematics of the source is rather stable, having a jet speed of less than 0.2\,$c$. The outer scale mas-scale jet is thought to be mis-aligned with an inclination angle of 65$^{\circ}$ estimated from downstream jet/ counter-jet ratios \citep[][]{fujita2017}. However, the high-energy observations, time-series analysis and  spectral energy distribution (SED) modeling, favor a smaller viewing angle, $\sim$20$^{\circ}$ \citep{abdo2009}.

The paper is structured as follows. Observations and data reduction are described  in Section \ref{sec:obs}. Results and analysis are given in Section \ref{sec:res}. 
Sections \ref{sec:discussion} and \ref{conclusions} presents the discussion and  conclusions, respectively.

\begin{table*}
	\centering
	\caption{Basic properties of each station.}
	\label{stations}
	\begin{tabular}{lcccc}
		\hline
		Station & Station code & Location & Effective diameter & SEFD \\
		 & & & [m] & [Jy] \\
		\hline
		Effelsberg & EF & Germany & 80 & 1500 \\
		Metsahovi & MH & Finland & 14 & 17500 \\
		Onsala & ON & Sweden & 20 & 5500 \\
		Plateau de Bure & PB &  France & 36.7 & 500 \\
		Pico Veleta & PV & Spain & 30 & 700 \\
		Yebes & YS & Spain & 40 & 1700 \\
		Green Bank Telescope & GB & United States & 100 & 100 \\
		VLBA & BR, FD, KP, LA, MK, NL, OV, PT & United States & 25 & 2000 \\
		\hline
		\hline
	\end{tabular}
\end{table*}

\begin{table*}
	\centering
	\caption{Summary of observations.}
	\label{observation}
	\begin{tabular}{lccc}
		\hline
		Epoch & Obs. time range (start/stop) & Participating stations & Beam size, position angle \\
		 & DOY/hh:mm [UT] & & [mas]$\times$[mas], [$^{\circ}$] \\
		\hline
		2008.36 & 130/00:15 ~ 132/01:35 & VLBA+MH+PB+PV+ON+EF & 0.125 $\times$ 0.053, -19.8 \\
		2011.35 & 127/06:00 ~ 128/01:50 & VLBA+MH+PB+PV+ON+EF & 0.144 $\times$ 0.048, -12.4 \\
		2013.74 & 270/00:00 ~ 271/13:35 & VLBA(-LA)+PV+ON+EF+YS & 0.129 $\times$ 0.048, -1.3 \\
		2014.40 & 145/12:30 ~ 147/20:50 & VLBA(-NL)+PV+ON+EF+YS & 0.123 $\times$ 0.047, -21.9 \\
		2014.73 & 268/21:15 ~ 272/12:20 & VLBA+PB+PV+EF+YS+GB & 0.179 $\times$ 0.047, -23.4 \\
		2015.37 & 136/12:30 ~ 136/20:50 & VLBA+ON+EF+GB & 0.155 $\times$ 0.056, -16.2 \\
		\hline
		\hline
	\end{tabular}
\end{table*}

\section{Observations and data reduction}\label{sec:obs}

\begin{figure}
    \centering
    \includegraphics[width=0.9\linewidth]{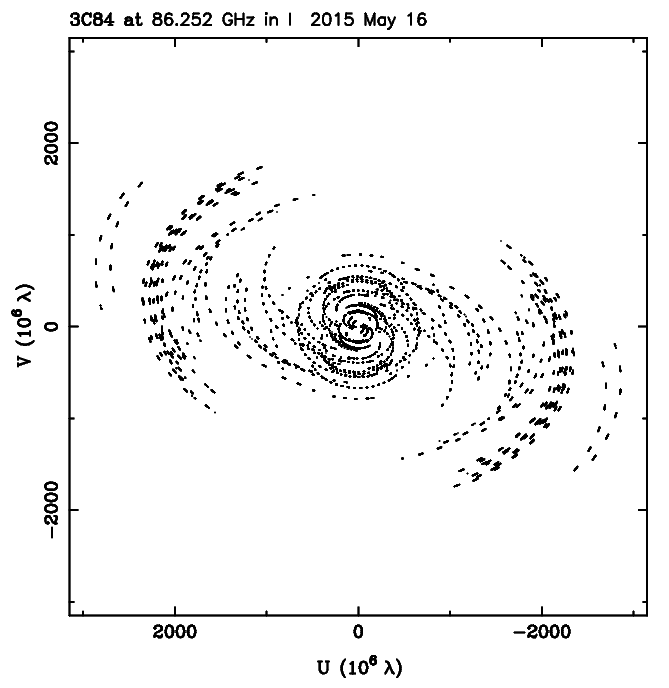}
    \caption[]{An example of \textit{uv}-coverage of GMVA 86 GHz observation of 3C\,84. This specific \textit{uv}-coverage is from the 2015 epoch.}
    \label{uv_cov}
\end{figure}

We used  GMVA observations at 3.5\,mm (86\,GHz) to study the central nucleus of 3C~84. 
The GMVA operates with 6 European antennas (Effelsberg, Onsala, Mets{\"a}hovi, Pico Veleta, Plateau de Bure, and Yebes) and 8 Very Long Baseline Array (VLBA) antennas located in the United States. In 2015, the 100-m Green Bank telescope, located in the United States, also participated in the observations. The basic properties of each telescope are listed in Table \ref{stations}. To ensure good \textit{uv}-coverage, all observations were conducted with a nearly full-track, with approximately 12 hours observing time per observation. Each observation consisted of many 5--7 min long VLBI scans; 
with scheduled gaps in-between, which were used for calibration and antenna pointing. The summary of observations is listed in Table \ref{observation}. Figure \ref{uv_cov} shows the typical \textit{uv}-coverage of our observation. 
The correlation was performed with the DiFX software correlator at the Max Planck Institute for Radio Astronomy, Bonn, Germany \citep{deller2011}. 
The  data set comprises 8 years of observations from 2008 to 2015.

Standard tasks from the NRAO Astronomical Image Processing System (AIPS) software package were used to analyse and calibrate the observed visibility data. The data were fringe fitted and calibrated in the usual manner, for details we refer to \citet{hodgson2017}. We CLEANed \citep{hogbom1974} the image with the Caltech Difmap software package \citep{shepherd1994}. Once the CLEANed images were made, we fit circular Gaussian model components to the maps. The uncertainties for each model component were calculated from extracted values of total flux density, peak intensity, radial distance from map center, position angle and size \citep[e.g.][]{lee2008}.

Large uncertainties in the flux density calibration are a known source of flux density error for the GMVA. \citet{kim2019} proposed a method to scale the visibility amplitudes of the GMVA observations from extrapolated lower-frequency total intensity observations, and suggested a antenna based scaling factor in the range of $\sim$2--2.5. Since our results are not sensitive to the absolute scaling of the amplitudes, we decided not to apply such amplitude corrections, but rely on the amplitude self-calibration method, which is implemented in Difmap. It is important to note that the flux densities presented in this paper may be underestimated by this factor.

\section{Results}\label{sec:res}

\subsection{Morphology}\label{sec:morph}

\begin{figure*}
\centering
\includegraphics[width=84mm]{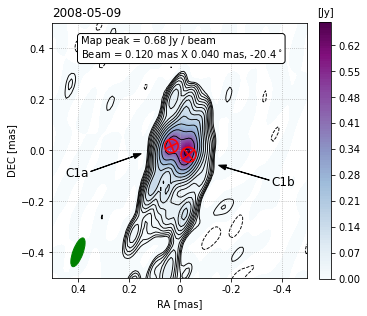}\hskip5mm
\includegraphics[width=84mm]{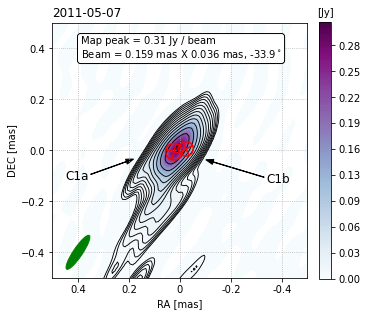} \
\includegraphics[width=84mm]{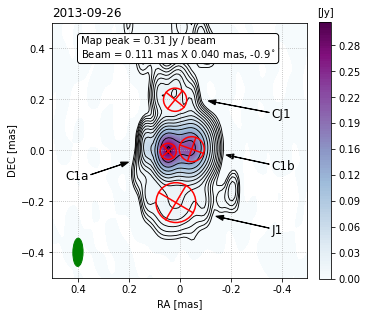}\hskip5mm
\includegraphics[width=84mm]{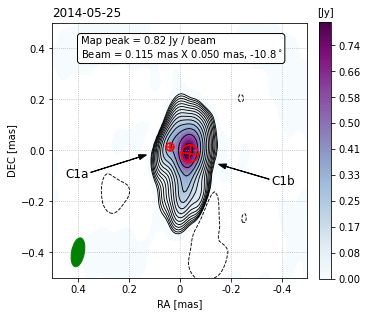} \
\includegraphics[width=84mm]{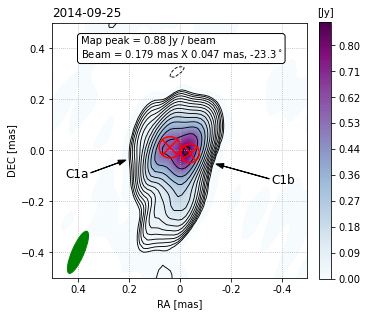}\hskip5mm
\includegraphics[width=84mm]{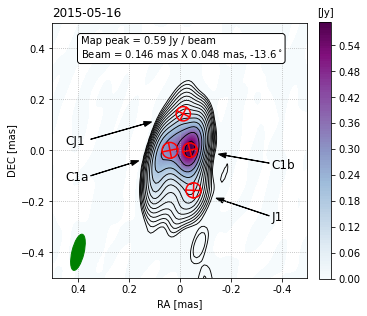} \
\caption{86 GHz GMVA maps of 3C\,84 from 2008.36 to 2015.37. The brightness distribution in the core region is fitted by up to 4 Gaussian components, with red circles indicating their positions. Green ellipses show the beam. The lowest contour level is 1\% of the map peak and increases by factors of $\sqrt{2}$.} 
\label{3c84_all}
\end{figure*}

Figure \ref{3c84_all} shows the GMVA 86 GHz VLBI maps of 3C\,84 at six different observing epochs.
Red circles represent the fitted model components, and the central nucleus. The C1  region, is distributed within a $\sim$0.3 mas radius of the map center. The  C1 region is resolved into at least two Gaussian components oriented in roughly east-west orientation, with some additional structures detected in a few cases. 
The components are labeled C1a (east) and C1b (west). We  calculated the brightness temperature of each model component using:
\begin{equation}
    \label{tbobs}
    T_{\rm B}^{\rm obs}=1.22\times 10^{12}\times \frac{S\left ( 1+z \right )}{r^{2}\nu ^{2}} ~~~ {\rm K},
\end{equation} 
where $S$ is the flux density in Jy, $r$ is the FWHM of the circular Gaussian model in mas, $z$ is the redshift and $\nu$ is the observing frequency in GHz \citep{lico2016}. On average, the brightness temperature of C1b (4.7$\times 10^{10}$ K) is higher than the brightness temperature of C1a (2$\times 10^{10}$ K), except in 2011.35, and 2013.74.

Interestingly, the jet is divided into two lanes, showing an inverted ``U'' shape at $\sim$0.15~mas south from the center.
This ``limb-brightened'' jet structure was also detected in the 7\,mm map shown in \citet{nagai2014} and  in the 22 GHz RadioAstron Space-VLBI maps in \citet{giovannini2018}.
The estimated separation between C1a and C1b is  $\sim$0.07~mas, corresponding to a separation of $\sim$770\,$R_g$. 
The components exhibit no motion, i.e., are quasi-stationary between 2008 to 2015. The derived upper limit of the motion is  25.1\,$\mu$as/year, i.e., 0.03\,$c$. We also find that the jet direction does not  significantly change over the course of the observations. For the upper limit of the motion, please refer to Figure \ref{epoch_err}, bottom panel. 

Additional structure is detected in the maps along the north-south direction. We label emission south of the C1 complex as ``J1'' and ``CJ1'', when it is north. Because of the low cadence of the data, no direct kinematic information can be determined for these components.

\subsection{Relationship between C1a and C1b}\label{sec:relation}
\begin{figure}
    \centering
    \includegraphics[width=\linewidth]{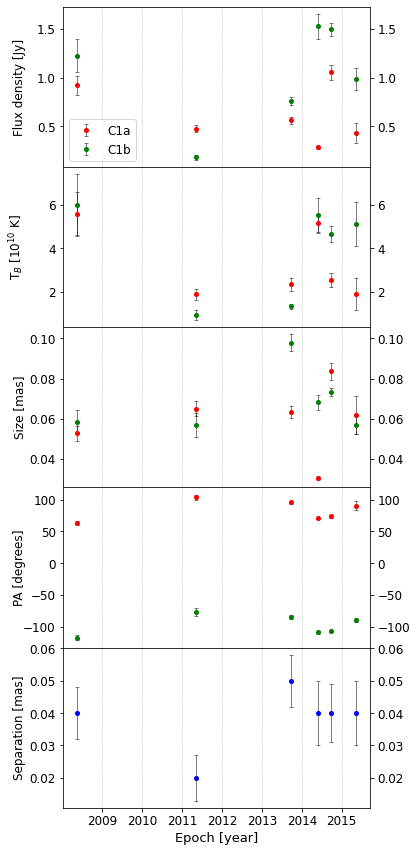}
    \caption[]{Flux density, brightness temperature, position angle and size of C1a (red) and C1b (green) over time. Separation (blue) is the distance from the midpoint between C1a and C1b.}
    \label{epoch_err}
\end{figure}

The detailed fit parameters for the Gaussian model components from 2008 onwards are summarized in Table \ref{mod_fits}. The values in col. 4 (`separation') represent the distance from the reference point. Since the reference point was chosen as the midpoint between C1a and C1b, the values are the same for C1a and C1b. The distance between C1a and C1b would be twice the value of the separation. We calculated the uncertainties of the parameters of each component following the approximation in \citet{fomalont1999}, when they are larger than 1/5 of the beam size. Motivated by a possible geometric physical origin (e.g. precession, jet rotation), we searched for possible trends or correlations between component flux densities, sizes and position angles. We did not find any significant trend in any parameter over time for both components C1a and C1b. Figure \ref{epoch_err} summarizes the results. 

\begin{figure}
    \centering
    \includegraphics[width=\linewidth]{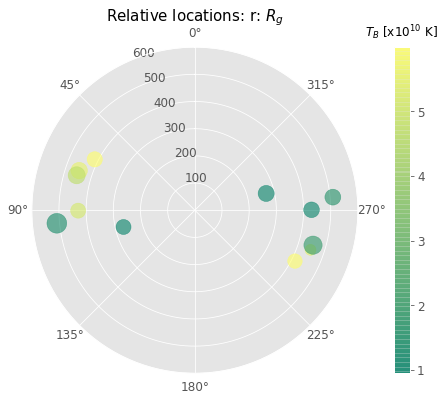}
    \caption[C1polar]{The relative location of C1a (group of the left) and C1b displayed in polar coordinates (separation in units of $R_{g}$, and position angle). The color coding marks the brightness temperature.}
    \label{C1polar}
\end{figure}

Taking the midpoints of C1a and C1b as the reference point, we plotted the brightness temperature and the sizes of C1a and C1b in a polar map in Figure \ref{C1polar}. The relative separation between C1a and C1b is smaller than the width of the jet, i.e. the distance between the edges of the jet that are seen in the 7\,mm maps \citep{nagai2014}. For C1a, there is a trend towards the higher brightness temperature in a northerly direction, while C1b shows a similar trend, except the higher brightness temperature trend to the south. In general, the C1a components appear smaller than C1b components. There is considerable variability in the brightness temperatures of C1a and C1b individually, with C1a varying by $\sim$3 times and C1b varying by a factor of $\sim$6 times. 

\begin{figure*}
    \centering
    \includegraphics[width=0.8\textwidth]{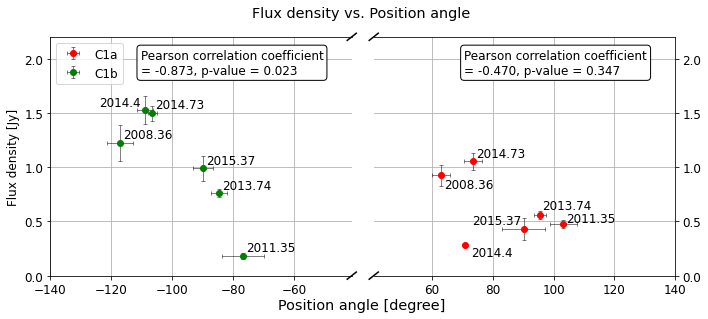}\\
    \includegraphics[width=0.8\textwidth]{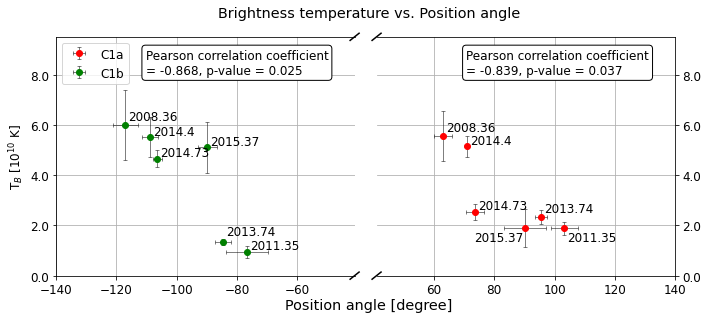}
    \caption[]{Flux density (top panel) and brightness temperature (bottom panel) of C1a (red) and C1b (green) as a function of position angle. The Pearson correlation coefficient with p-value is shown in the box at top-right of each plot.}
    \label{c1ab_cor}
\end{figure*}

In order to estimate the significance of the observed trends, we computed the Pearson correlation coefficients for the flux density and brightness temperature for C1a and  C1b. We found a significant (p $<$ 0.05) correlation in $T_{\rm B}$ and flux density with position angle (Figure \ref{c1ab_cor}). 
Applying a constant scaling factor we discussed in Section \ref{sec:obs} would not affect our results, however, a time variable scaling factor could affect our results. Since we do not expect the compactness of the source to change significantly as a function of time, we do not think it is likely that there is a time variable scaling factor. While the correlation appears significant, we caution that there are a small number of observations and more observations should be made in order to confirm the observed correlation.

\begin{table*}
\centering
\caption{Physical parameters of each model component using the midpoint between C1a and C1b of each epoch as the reference. ``Separation'' in the 4th column means the distance from the reference point.}

        \begin{tabular}{lcccccc}
            \hline
            ID & Epoch & Flux  density    & Separation & PA & Size & T$_B$ \\
		   &      &  [Jy] & [$\mu$as] & [degrees] & [$\mu$as] & [$\times 10^{10}$K]   \\
		    \hline
		    C1a & 2008.36 & 0.92 $\pm$ 0.10 & \multirow{2}{*}{38 $\pm$ 8} & 63.1 $\pm$ 2.9 & 53 $\pm$ 4 & 5.56 $\pm$ 1.00 \\
            C1b & 2008.36 & 1.23 $\pm$ 0.17 &  & -117.0 $\pm$ 4.3 & 59 $\pm$ 6 & 5.99 $\pm$ 1.40 \\\hline
            C1a & 2011.35 & 0.47 $\pm$ 0.03 & \multirow{2}{*}{25 $\pm$ 7} & 103.2 $\pm$ 4.4 & 65 $\pm$ 4 & 1.88 $\pm$ 0.25 \\
            C1b & 2011.35 & 0.18 $\pm$ 0.03 &  & -76.8 $\pm$ 6.9 & 57 $\pm$ 6 & 0.95 $\pm$ 0.24 \\\hline
            C1a & 2013.74 & 0.56 $\pm$ 0.04 & \multirow{2}{*}{46 $\pm$ 8} & 95.4 $\pm$ 2.0 & 63 $\pm$ 3 & 2.34 $\pm$ 0.28 \\
            C1b & 2013.74 & 0.76 $\pm$ 0.04 &  & -84.6 $\pm$ 2.6 & 98 $\pm$ 4 & 1.33 $\pm$ 0.13 \\
            J1 & 2013.74 & 0.13 $\pm$ 0.02 & 206 $\pm$ 10 & 175.6 $\pm$ 2.8 & 156 $\pm$ 20 & 0.09 $\pm$ 0.02 \\
            CJ1 & 2013.74 & 0.06 $\pm$ 0.01 & 199 $\pm$ 8 & 5.5 $\pm$ 1.7 & 91 $\pm$ 12 & 0.12 $\pm$ 0.04 \\\hline
            C1a & 2014.4 & 0.28 $\pm$ 0.01 & \multirow{2}{*}{41 $\pm$ 10} & 71.1 $\pm$ 0.6 & 30 $\pm$ 1 & 5.14 $\pm$ 0.42 \\
            C1b & 2014.4 & 1.53 $\pm$ 0.13 &  & -108.9 $\pm$ 2.7 & 68 $\pm$ 4 & 5.51 $\pm$ 0.78 \\\hline
            C1a & 2014.73 & 1.06 $\pm$ 0.08 & \multirow{2}{*}{41 $\pm$ 9} & 73.5 $\pm$ 2.9 & 84 $\pm$ 4 & 2.53 $\pm$ 0.32 \\
            C1b & 2014.73 & 1.50 $\pm$ 0.07 &  & -106.5 $\pm$ 1.5 & 73 $\pm$ 2 & 4.66 $\pm$ 0.35 \\\hline
            C1a & 2015.37 & 0.43 $\pm$ 0.10 & \multirow{2}{*}{39 $\pm$ 10} & 90.2 $\pm$ 7.0 & 62 $\pm$ 10 & 1.89 $\pm$ 0.74 \\
            C1b & 2015.37 & 0.99 $\pm$ 0.12 &  & -89.8 $\pm$ 3.2 & 57 $\pm$ 4 & 5.11 $\pm$ 1.02 \\
            J1 & 2015.37 & 0.15 $\pm$ 0.03 & 166 $\pm$ 10 & -161.3 $\pm$ 1.4 & 59 $\pm$ 8 & 0.73 $\pm$ 0.25 \\
            CJ1 & 2015.37 & 0.04 $\pm$ 0.02 & 144 $\pm$ 10 & -5.2 $\pm$ 3.0 & 56 $\pm$ 15 & 0.22 $\pm$ 0.02 \\\hline\hline
        \end{tabular}

    \label{mod_fits}
\end{table*}

\subsection{Stacked map}\label{sec:stacked_map}

\begin{figure*}
    \centering
    \includegraphics[width=\textwidth]{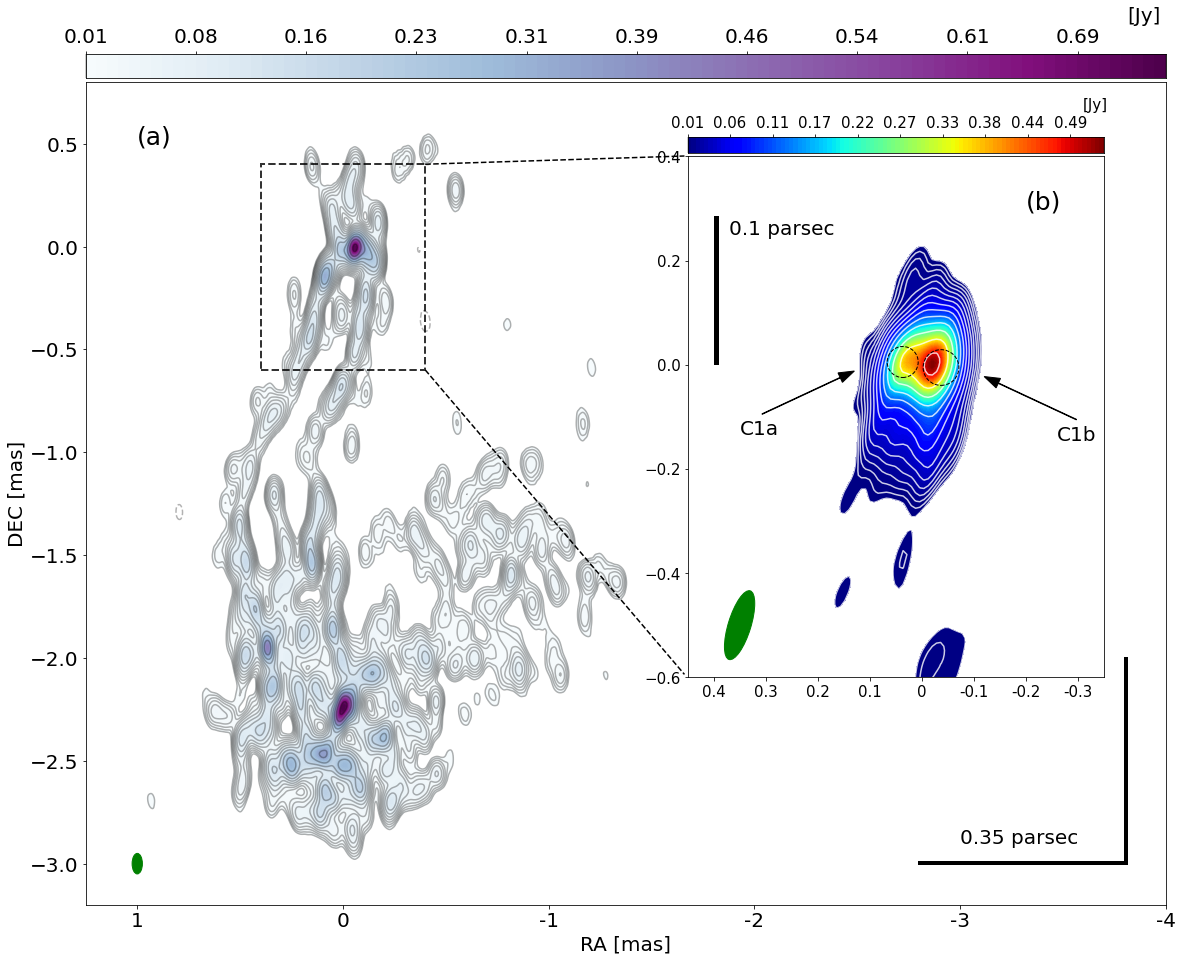}
    \caption[stacked]{(a) 22 GHz RadioAstron image. The map peak is 0.75 Jy/beam with the beam size of 0.1 $\times$ 0.05 mas, 0$^\circ$. The noise level is 1.4 mJy/beam \citep{giovannini2018}. (b) 86 GHz GMVA stacked map. Before stacking, all 6 epochs were restored with the averaged beam of 0.138 $\times$ 0.0435 mas, -17.15$^\circ$. The midpoint of C1a and C1b was used for the reference point. Dashed circles represent C1a and C1b with the averaged position and the averaged size. The map peak is 0.53 Jy/beam and the noise level is 1.57 mJy/beam.
    The lowest contour level is 1\% of the map peak and increases by factors of $\sqrt{2}$ for both RadioAstron and GMVA stacked map. The green ellipse at bottom-left corner of each map indicates the synthesized beam.}
    \label{stacked}
\end{figure*}

Because VLBI imaging can be subjective, in order to be more confident in our results, we stacked all 6 epochs without specific weights. In order to stack the maps, we restored each map with the same averaged beam (0.138 [mas] $\times$ 0.0435 [mas], -17.15$^\circ$) and added them together and then divided by 6. We also have to choose a reference point to align the maps on. We tried three options, alignment i) on C1a, ii) C1b and iii) the midpoint between C1a and C1b. In all three cases, the stacked map clearly shows an east-west stretched nuclear region  (Figure \ref{stacked}). In the south, a prominent double-railed structure is visible, which most likely connects to the limb-brightened jet structure further down (beyond C3). One could argue that the structure might be triple-railed rather than double, as another (more marginal) ridge is seen between two peaks in the C1 complex. However this cannot be unambiguously determined by the angular resolution of this data set. Similar but fainter structure is also seen to the north. In the stacked map, C1b is brighter than C1a.

\section{Discussion}\label{sec:discussion}

\subsection{What is the reference point?}

In Figure \ref{c1ab_cor}, we present the correlation between physical parameters of C1a and C1b, which appears to be significant, albeit based only on a small number of observations over a limited period of time. We also note a better correlation between flux density (or $T_{\rm B}$ ) and position angle for C1b, than for C1a.

The correlations between flux density (or $T_{\rm B}$) and PA does not depend on the choice of a reference point. However, for the visualization in the polar diagram of Figure \ref{C1polar}, we had to define a reference point, which we set arbitrarily as the midpoint between C1a and C1b. This choice was motivated by the RadioAstron image, suggesting that C1a and C1b may correspond to the outer boundary of a transversely resolved jet.

\subsection{The physical condition of the jet base}\label{sec:plasma_physics}

In Section \ref{sec:relation} and seen in Figure \ref{C1polar}, we found significant (p $<$ 0.05) correlation between the brightness temperature and the position angle in both C1a and C1b. The brightness temperatures vary by factors up to about 6 (in C1b) and show systematically lower values toward larger position angles. This behaviour can be understood as the effect of emitters moving on a helical path around the outer layer of the jet.

If the emitter is moving at relativistic speeds, observer frame and emitter frame brightness temperatures, $T_{\rm B}^{\rm obs}$ and $T_{\rm B}^{\rm em}$ in K, respectively, are related via the Doppler-factor as follows:

\begin{equation}
    \label{TB-boosted}
    T_{\rm B}^{\rm obs} = \delta T_{\rm B}^{\rm em}
\end{equation}
where $\delta$ is the Doppler factor
\begin{equation}
    \label{delta}
    \delta = \left[ \Gamma (1 - \beta\cos\theta) \right]^{-1}
\end{equation}
with $\beta$ being the speed in units of speed of light $c$, $\Gamma$ being the Lorentz factor of the emitter, and $\theta$ being the angle between emitter path and line of sight. Equation~\ref{TB-boosted} shows that a variation of brightness temperature due to Doppler boosting by a factor of six requires an equal variation of the Doppler factor. Even in the extreme case that $\theta$ varies from $0^{\circ}$ to $180^{\circ}$ -- which is actually excluded because C1a corresponds to one side of the jet only -- we require $\beta\geq0.7$. Such high speeds are, arguably, at odds with the observed proper motions of the jet components of 3C\,84 that are well below $c$ and imply $\beta\ll1$. 

A further hint toward the intrinsic jet kinematics is provided by the observed brightness temperatures themselves. Within errors, all observed values obey $T_{\rm B}^{\rm obs} \leq 2\times10^{11}$~K, in agreement with the physical limit given by the equipartition of energy between radiating electrons and magnetic field \citep{readhead1994, singal2009}. It should be noted that even if we applied a correction factor of $\sim$2 to the flux densities (see Section \ref{sec:obs}), the observed brightness temperatures would still be $T_{\rm B}^{\rm obs} \leq 2\times10^{11}$~K and therefore consistent with being in equipartition. The fact that the maximum \textit{observed} brightness temperature respects the equipartition limit for the \textit{emitter frame} brightness temperature suggests $\delta\approx1$ and thus $\beta\ll1$.

The second physical mechanism to consider is the intrinsic evolution of the jet plasma after its ejection from the central engine. Making the standard assumption that all radiation we observe from the jet is synchrotron radiation, any emitter will be subject to synchrotron cooling while moving outward. The fact that our approximately annual sampling does not show a correlation of flux density with time (see Figure \ref{epoch_err}) implies that the synchrotron cooling timescale is well below one year (we do not have the cadence to probe shorter timescales). Instead of a single emitter cooling down over several years, we rather likely observe multiple individual emitters emerging from the central engine and cooling down rapidly. The cooling timescale (in the observer frame) is related to the physical parameters of the jet like
\begin{equation}
    \label{tau-cool}
    \tau_{\rm cool} = 7.74 \left[\frac{\delta}{1+z}\right]^{-1} B^{-2} \gamma^{-1} ~~~ {\rm seconds}
\end{equation}
where $B$ is the magnetic field strength in T, $\gamma$ is the (averaged) Lorentz factor of the gyrating electrons, and $z$ is the redshift. As discussed above, we may assume $\delta\approx1$; we also know $z=0.0176$. The $\sim$6 monthly cadence of the observations gives an upper limit of $\sim$3 months to the cooling timescale. Cooling timescales of $\sim$3 months would be consistent with $B\approx10~\mu$T and $\gamma\approx10\,000$; which are typical values for AGN. However, higher cadence observations could reveal shorter timescales and correspondingly higher magnetic fields.

This is potentially interesting since \citet{hodgson2018b} also showed that the C1 region could exhibit blazar-like mm-wave spectral behaviour, despite likely having a relatively low Lorentz factor. This is surprising because blazar emission is thought to mostly originate in downstream recollimation shocks rather than so close to the jet base \citep{jorstad2005, hodgson2017}.

In the case of C1a, a combination of synchrotron cooling and expansion of the plasma flow can explain the observed correlations of flux density and brightness temperature with position angle. The plasma flow experiences synchrotron cooling, leading to a decrease in flux density by a factor of two along its way. With brightness temperature being, effectively, flux density per angular area, an increase of the (projected) emitter diameter by a factor of $\sqrt{2}$ and thus of the area by a factor of two along the trajectory makes the brightness temperature decrease by a factor of about four. Overall, we were probably lucky to observe systematic trends at all: assuming emission from multiple emitters, substantial variations in the initial flux densities and/or brightness temperatures from source to source may easily mask intrinsic correlations.

\subsection{Are C1a and C1b the true jet base?}\label{sec:jetbase}

\noindent We can potentially test the scenario that the C1 complex is a BP jet rooted in the accretion disk. In order to test this, we estimated the size of the broad line region (BLR) from H-$\beta$ lines. The broad line region, $R_{\rm BLR}$, can be estimated using a tight empirical correlation between the size of the BLR and the monochromatic continuum luminosity \citep{bentz2013}:

\begin{equation}
    R_{\rm BLR} = (33.651 \pm 1.074) \left( \frac{L_{5100}}{10^{44}\ erg\ s^{-1}} \right) ^{(0.533 \pm 0.034)}  \textrm{lt-days} .
\end{equation}\

\noindent L$_{5100}$ can also be derived from a correlation between the monochromatic continuum luminosity and the line flux density \citep{greene2005}:

\begin{equation}
    L_{\rm H\beta} = (1.425 \pm 0.007) \times 10^{42} \left( \frac{L_{5100}}{10^{44}\ erg\ s^{-1}} \right) ^{(1.133 \pm 0.005)}  \textrm{erg  s$^{-1}$} .
\end{equation}\

\noindent Then the size of the BLR can be estimated from L$_{\rm H\beta}$:

\begin{equation}
    R_{\rm BLR} = 28.487 \times \left( \frac{L_{\rm H\beta}}{10^{42}\ erg\ s^{-1}} \right) ^{0.470} \textrm{lt-day}  .
\end{equation}\

\noindent Using L$_{\rm H\beta}$ taken from \citet{punsly2018}, we find that the size of the BLR is 7.7--10.8 light-days, which is already roughly a factor of 3 smaller than the separation of C1a and C1b ($\sim$30 light days). Since the accretion disk is highly unlikely to be larger than the BLR, we therefore consider it unlikely that the C1 complex represents a BP jet anchored in the accretion disk. We note with interest however, that the estimated size of the BLR is similar to the separation of C1a and C1b.

\subsection{Black Hole location}

\begin{figure*}
    \hskip0mm
    \includegraphics[width=50mm]{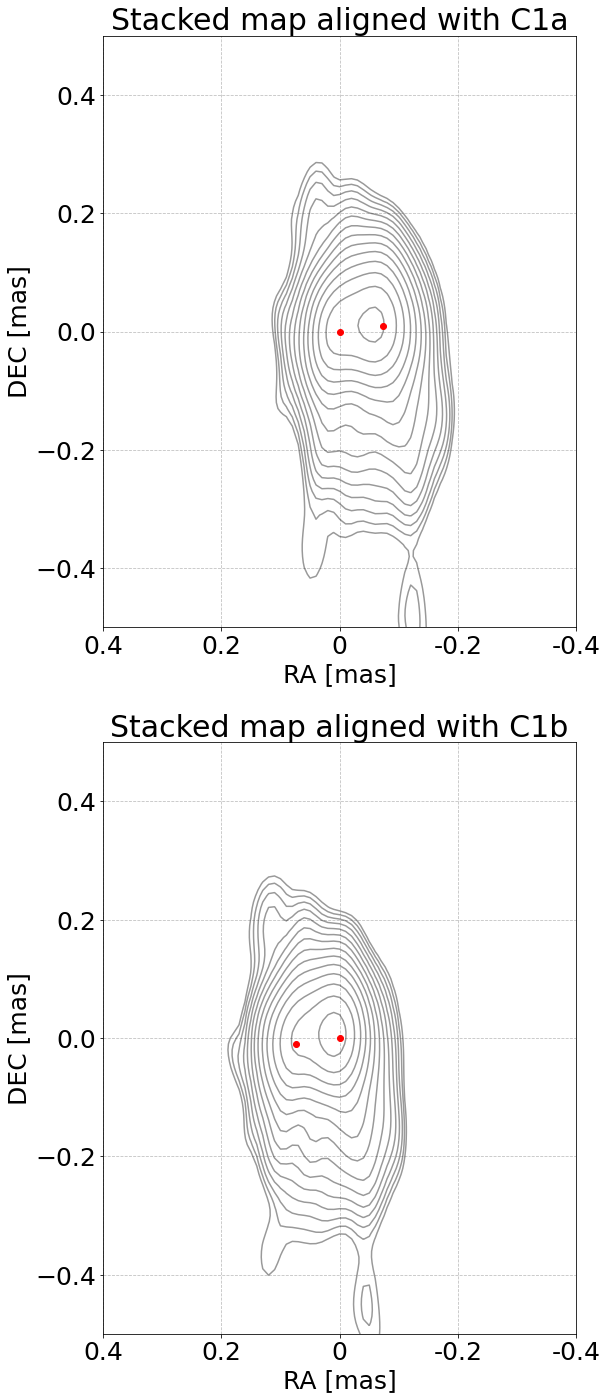}\hskip10mm
    \includegraphics[width=100mm]{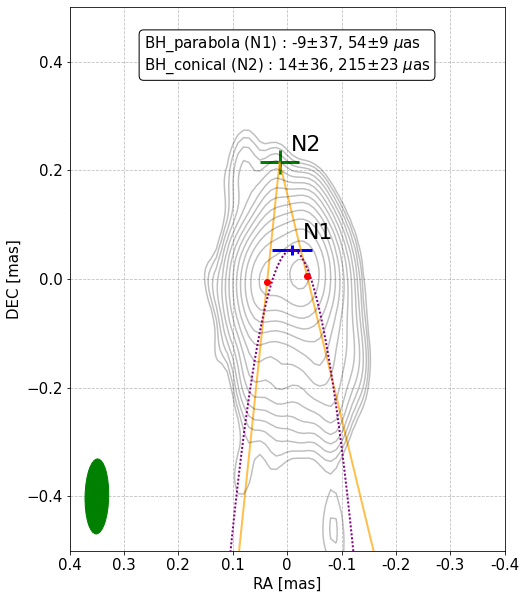}
    \caption{\textit{Top left} : Stacked map aligned in the position of component} C1a. \textit{Bottom right} : Stacked map aligned on C1b. \textit{Right} : Estimation of Black Hole location using conical and parabolic fits to the transverse jet profile. For this calculation the maps were aligned on the mid-point between C1a and C1b. For better display and ease fitting, the maps are rotated by 16$^{\circ}$ counter-clockwise. The estimated locations of each fit with error are represented as blue cross (N1, parabolic), and green cross (N2, conical). Red dots indicate the average position of C1a and C1b. The green ellipse at bottom-left corner indicates the synthesized beam. The beam size, map peak, noise level and the contour levels are the same as Figure \ref{stacked}.
    \label{bh_loc}
\end{figure*}

If C1a and C1b are not the jet base, the SMBH is likely located further upstream of this emission. We can therefore place limits on the SMBH location using a transverse jet width profile comparison with the stacked map (Figure \ref{stacked}). We first rotated the map by $16^{\circ}$ to aligne the C1 complex to the horizontal axis. We then sliced the map horizontally, to obtain 1 dimensional flux density profiles along this axis. 
To measure the jet ridge in the nuclear area, we found the second derivative of each slice. We applied both conical (orange line) and parabolic (purple dashed line) fits with these points. We also applied extra weight on the position of C1a and C1b (red dots), to make sure the fits pass through the C1 complex. The result is displayed in Figure \ref{bh_loc}. 

The offset between the Black Hole position and the jet apex is unknown, but mm-VLBI images on 2 sided radio jets indicate that it may be smaller than a few hundred Schwarzschild radii \citep[e.g.,][]{baczko2016, boccardi2016, boccardi2017, blandford2019}.
In the case of 3C\,84 which is discussed in this paper an offset of $\leq200$\,$R_S$ would correspond to an angular separation of only $\sim$18\,$\mu$as, which is smaller than the angular resolution of the available GMVA maps. Since the effect is tiny and for simplicity in the following discussion, we tentatively associate the location of the jet apex and the Black Hole position to be identical with each other and postpone a more rigorous analysis of possible a displacement of the jet origin from the Black Hole to future work, when VLBI images with higher angular resolution become available.

In order to test these scenarios, we calculated the reduced $\chi^{2}$ for both the conical and parabolic fits. The parabolic fit produced the lower reduced $\chi^{2}$ of 0.7, but as it is less than 1, it could be overfitting the data. The conical fit also fits the data well, with a reduced $\chi^{2}$ of 1.07. We therefore cannot make a strong statement on which fit the data prefers. 

The estimated location of the Black Hole, or the jet apex, is between $\sim$54 and $\sim$215\,$\mu$as upstream of the 3\,mm core region. We labeled them as N1 (for the parabolic fit) and N2 (for the conical fit). The projected distance adopting the Black Hole mass of 3.2 $\times 10^8 M_\odot$ is $\sim$590\,$R_g$ and $\sim$2360\,$R_g$, respectively. 

The uncertainties of the estimated Black Hole location mainly come from the image alignment. We have used the midpoint of C1a and C1b as the reference point throughout this paper, however we cannot rule out the possibilities of having different reference points. We therefore created three stacked maps, referencing against C1a, C1b and the midpoint. We have fitted N1 and N2 on all three maps, and obtained the standard deviation of the position of N1 and N2. In addition, we applied the position error described in \citet{fomalont1999} onto the weighted positions while fitting, and obtained the possible range of N1 and N2. For N1, the position uncertainty is at the few $\mu$as level. For N2, however, very small variations yield large uncertainty in the north-south direction. We took the larger values from both methods for our estimations. The estimated positions with errors for N1 and N2 are $9\pm37,54\pm9$ $\mu$as and $14\pm36,215\pm23$ $\mu$as, where (0,0) is the midpoint between C1a and C1b as used in Figure \ref{bh_loc}. 

\subsection{Jet / Counter-jet and viewing angle}\label{sec:jcj}

In \citet{abdo2009}, a viewing angle from the spectral energy distribution (SED) fitting of $\gamma$-ray data was derived to be $\sim$25$^{\circ}$, which is in tension with the results from \citet{fujita2017}, where a viewing angle of $\sim$65$^{\circ}\pm16^{\circ}$ was derived from a jet/counter-jet analysis performed further south and downstream from the C1 region discussed in this paper. \citet{hodgson2018b} showed that during the period that the \citet{abdo2009} analysis was conducted, the $\gamma$-ray emission could have possibly been occurring in the C1 region. This suggests the possibility that the viewing angle varies between nuclear region and the outer more extended jet. We therefore have tried to test this scenario.  

If we assume that the jet and counter-jet are intrinsically symmetric, the jet/counter-jet ratio $R$ is related to the speed of the jet $\beta$ in $c$ and the viewing angle $\theta$ by the expression:

\begin{equation}
    R=\left(\frac{1+\beta_{\rm }\cos(\theta)}{1-\beta_{\rm }\cos(\theta)}\right)^{\mu}
\end{equation}\

\noindent where $\mu$ is either $2-\alpha$ or $3-\alpha$ depending if the jet is continuous or a single component \\citep{blandford1979}. In the case of 3C\,84,  we see a continuous jet and assume a flat spectrum in the region \citep{wajima2020} and therefore use $\mu = 2$. Since we do not have kinematic information in the direct vicinity of the C1 region from these observations, we can assume viewing angles from the literature and derive an estimate of $\beta$ for the C1 region. Observations of the kinematics closer to the C1 region have speeds of $\beta_{\rm app}\sim$0.1--0.2 \citep[e.g.,][]{krichbaum1992, dhawan1998, suzuki2012}, and also more recently a value of $\beta_{\rm app}=0.086$ was reported by \citet{punsly2021}. We note that higher speeds (e.g. \citet{hiura2018} : 0.27\,$c$, \citet{lister2009} : 0.31\,$c$ and \citet{suzuki2012} : 0.47\,$c$) have been reported for the C3 region.

\begin{figure*}
    \centering
    \includegraphics[width=\linewidth]{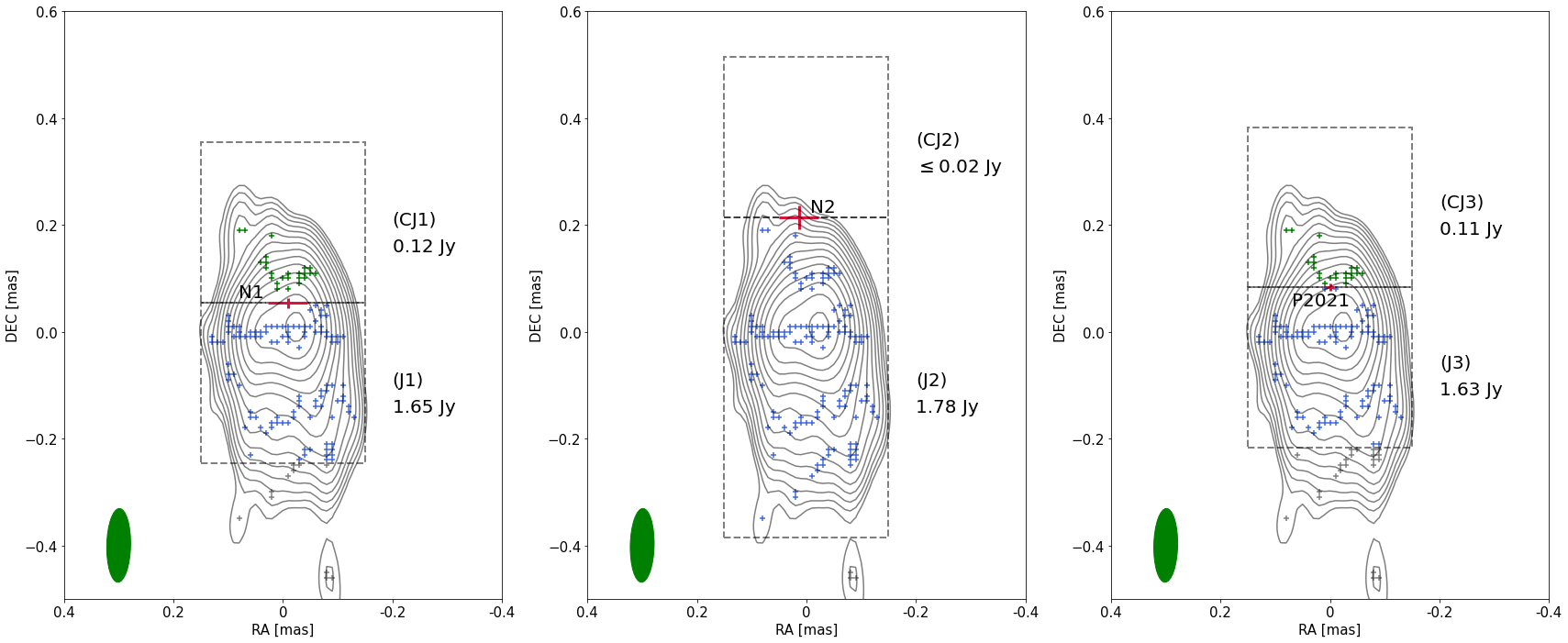}
    \caption[jetcjet2]{This Figure shows the results of the jet/counter-jet analysis as described in Section \ref{sec:jcj}. Green crosses are CLEAN components counted on the counter-jet side, blue crosses are CLEAN components counted on the jet side and grey crosses are CLEAN components that were not included in the analysis. The dashed black boxes have a size of 0.3 mas $\times$ 0.3 mas and the CLEAN components within them are summed in order to perform the analysis. The dividing lines are, from the left, N1, N2 (as described in Section \ref{bh_loc} and Figure \ref{bh_loc}) and P2021, which is the approximate location of the jet apex as measured by \citet{paraschos2021}. The flux density of summed CLEAN components in each box is indicated next to the boxes. In the case of N2, the noise level was used as an upper limit on the flux density of the counter-jet. The counter-jet box is twice as large as plotted in the image. The green ellipse indicates the synthesized beam. The beam size, map peak, noise level and the contour levels are the same as Figure \ref{stacked}.}
    \label{jetcjet_err}
\end{figure*}

We used the parabolic (N1) and conical (N2) jet apex locations from Figure \ref{bh_loc} as the dividing lines. From this we obtain a jet apex position of 54$\pm$9\,$\mu$as for N1, and 215$\pm$23\,$\mu$as for N2. From an independent core-shift analysis based on a single epoch image (May 2015), \citet{paraschos2021} recently reported a jet apex location 83$\pm$7\,$\mu$as north of the C1ab complex. This is close to the position, which we derive for N1. In Figure \ref{jetcjet_err}, we show maps for the 3 scenarios (N1, N2 and P2021).

We decomposed the stacked map -- which is a superposition of six CLEANed maps restored with the same clean beam (Section~\ref{sec:stacked_map}) -- into individual components by CLEANing it (again) using the known clean beam as dirty beam and storing the resulting component map. Effectively, we use the CLEAN procedure as a component detection algorithm.\footnote{Whereas this approach is unusual in radio astronomy, it is well-known in optical astronomy. An example is provided by the famous \href{http://www.star.bris.ac.uk/~mbt/daophot/}{DAOPHOT} software package \citep{stetson1997} which uses a CLEAN-like procedure (\texttt{SUBSTAR}) to subtract and deblend stars in images of crowded stellar fields.}
We identified 147 CLEAN components in total and calculated the flux density in the stacked map by summing the clean components.

The green and blue crosses in the dashed box represent the clean components as shown in Figure \ref{jetcjet_err}. The green ellipse at bottom-left corner is the averaged beam from 6 epochs, which is also used in the stacked map, but rotated by 16$^{\circ}$ counterclockwise along with the map.

In each map we summed the CLEAN components within two boxes north and south of each dividing line. On the jet side, we have labeled these boxes as J1, J2 and J3, and for the counter-jet side as CJ1, CJ2 and CJ3 respectively. For the flux determination in the jet and counter-jet regions, we used equally sized boxes, assuming that the relativistic length contraction is negligible (which is a reasonable assumption given the measured low jet speed). In the case N1 and P2021, an area of 0.3 mas width and 0.3 mas height was used. This corresponded to about 19 times the beam size. In the case of N2, it was necessary to double the height of the summed region. We calculated the jet/counter-jet ratio by summing the CLEAN components within each box, i.e., blue/green for J/CJ. However, we did not find any clean components in the case of N2. We therefore used the noise level ($\leq$1.1 mJy/beam) of the map, multiplied by the number of beams fitting inside the box (CJ2), as an upper limit of the counter-jet flux density. As J2 is twice as big as CJ2, we doubled the value of CJ2. The calculated jet/counter-jet ratio is 13.7 and 15 for N1 and P2021. In the case of N2, a lower limit of 42.7 was found.

\begin{figure*}
    \centering
    \includegraphics[width=0.8\textwidth]{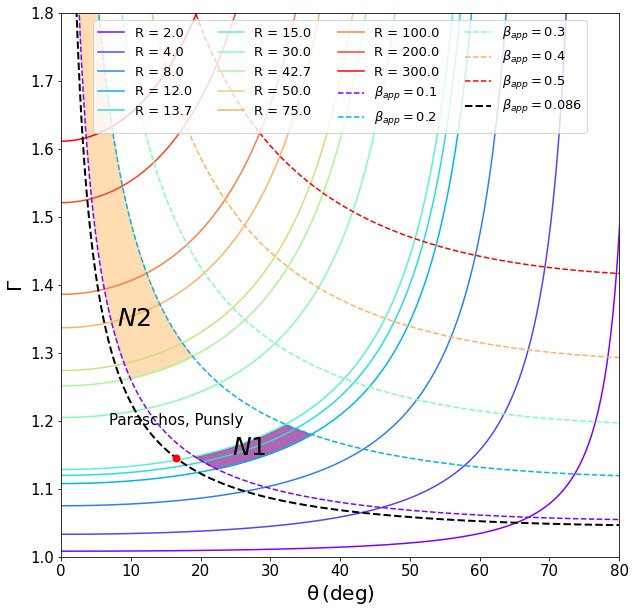}
    \caption[gammatheta]{The relationship between the Lorentz factor $\Gamma$, the apparent speed $\beta_{\rm app}$, the jet/counter-jet ratio $R$ and the viewing angle $\theta$ between the jet and our line of sight. If the jet base is located at N1, a viewing angle in the range of $\sim$20--35$^{\circ}$ is obtained (adopting an uncertainty of 15\% for R). For N2, we estimate an upper limit of the viewing angle of $\leq20^{\circ}$. The black dashed line represents $\beta_{\rm app} = 0.086$ from \citet{punsly2021}. The red dot represents the viewing angle if the result of the location of the jet apex from \citet{paraschos2021} is combined with the $\beta_{\rm app}$ results of \citet{punsly2021}.} 
    \label{gammatheta}
\end{figure*}

Figure \ref{gammatheta} shows the relationship between the Lorentz factor $\Gamma$, the apparent speed $\beta_{\rm app}$, jet/counter-jet ratio $R$ and the viewing angle $\theta$ between the jet and our line of sight.
Assuming $\beta_{\rm app}$ to be 0.1--0.2 from previous mm-VLBI imaging \citep[e.g.,][]{krichbaum1992, dhawan1998, suzuki2012}, and using $R$ = 13.7, with 15\% uncertainty for N1, the viewing angle $\theta$ is roughly $\sim$20--35$^{\circ}$ (N1 in Figure \ref{gammatheta}). 
For N2, the lower limit of $R\geq42.7$ gives the upper limit in the viewing angle of $\leq20^{\circ}$.

The viewing angle depends on the jet/counter-jet ratio, and the ratio is highly sensitive to the position of the reference point, which is assumed to be the jet apex, which we are also assuming to be also the location of the Black Hole. In previous section, we discussed that the potential Black Hole location could be between N1 and N2, and we can therefore can set the upper limit of the viewing angle to be $35^{\circ}$ from these observations.

A recent study by \citet{punsly2021} found $\beta_{\rm app}$ = 0.086$\pm$0.08 in the C1 region, which indicates a faster apparent jet speed than seen in our earlier observations yielding an upper limit of $\leq 0.03$\,c. (See Section \ref{sec:morph}).
The $\sim$3 times higher apparent speed may indicate jet acceleration or a slow decrease
of the jet viewing angle (e.g. caused by jet rotation). Since the measurements of  \citet{punsly2021} are closer in time than our measurement, we combine their speed measurement with the results from \citet{paraschos2021} to derive a jet viewing angle
of $17^{\circ}$.
In all cases, the viewing angle in the C1 region is less than about 35$^{\circ}$ and could be yet smaller than this limit. This is considerably smaller than the viewing angle of $65^{\circ}$ reported by \citet{fujita2017} in the downstream C3 region. We interpret this as tentative evidence for a variation of the jet viewing angle between the C1 and C3 region and therefore for a bent jet. Additionally, this provides indirect evidence for $\gamma$-ray emission originating in the C1 region (in addition to the C3 region), because it could explain the smaller viewing angle found in the $\gamma$-ray SED analysis of \citet{abdo2009}.

A small viewing angle would amplify variability (see Figure \ref{epoch_err}) and could explain the jet acceleration from $\leq$ 0.1--0.2\,$c$ near the core (C1) to 0.4-0.9\,$c$ further out (C3) \citep{hodgson2021}. The low apparent speeds could be explained if the viewing angle is smaller than the critical angle that maximizes the apparent speeds. This would lead to low observed values of $\beta_{\rm app}$ and Doppler factors of $\sim$ 1.5--2. Flux density variability studies at lower frequencies suggest a Doppler factor close to $\sim$1 in total intensity observations, which are also sensitive to the outer jet regions \citep[e.g.][]{hovatta09,liodakis17}. On the other hand, very recent observations in the publicly available Boston University 43\,GHz monitoring program \citep{jor17}, suggest a large and rapid change in the jet direction, which are more difficult to explain with a larger viewing angle.

It is known that 3C\,84 has free-free absorption features (\citep[e.g.,][]{walker2000, wajima2020}). It is possible that the presence of an unknown absorber could be interfering with our measurement of the jet/counter-jet ratio. In this case, the intrinsic (unabsorbed) jet/counter-jet ratio would likely be lower in reality and therefore prefer a larger angle solution. Higher cadence kinematic monitoring and accurate spectral image maps would be required to resolve the issue.

\subsection{Comparison with RadioAstron}

In the publication of \citet{giovannini2018}, the authors presented a space-VLBI map of 3C\,84 that was observed within a few days of our 2013 observation. In their map (reproduced in Figure \ref{stacked}, an E-W elongated structure is seen, and appears to be composed of three components. The central component is the brightest of these components. This is broadly consistent with the E-W elongation observed in the GMVA data presented here. In particular, the 2013.74 epoch has the most elongated structure seen in our observations. However, unlike the triple structure seen in the RadioAstron map, at 3\,mm, this structure is best fitted using 2 circular Gaussian components that we name C1a (east) and C1b (west), which are separated by $\sim$70\,$\mu$as. Additionally, beyond the central component, the space-VLBI map shows a limb brightened jet, while our maps typically show brighter centrally located emission. The differences are surprising, given that the space-VLBI observations are of higher resolution than ours. The recent work by \citet{paraschos2021} places the 3\,mm VLBI core 0.35 mas upstream of the 15 GHz core. 
So we consider it unlikely that the C1 region is the true jet base (see Section \ref{sec:jetbase}). Instead, it is likely that C1 is related to an oblique shock or otherwise transverse structure, which is located a bit downstream of the jet apex. Different opacities at the different frequencies then may cause the observed morphological differences between 22\,GHz and 86\,GHz. Another reasonable way to interpret the differences between the maps could be transverse jet stratification \citep[e.g.][]{ghisellini2005} with emission peaks at 22 and 86 GHz not being co-spatial but transversely displaced. Furthermore, under the jet stratification scenario, the broad-band SED modeling may also become more compatible with the VLBI observations \citep[e.g.][]{abdo2009}. 
While the current observations are unable to resolve this, future multi-wavelength observations which resolve the jet transversely will help to discriminate between the shock, the stratification and possible other scenario.

\section{Conclusions}\label{conclusions}

In this paper, we have analysed six high resolution (beam $<50 \mu$as) GMVA maps of the radio-galaxy 3C\,84 over the period 2008--2015. In its VLBI core region we see a persistent transverse nuclear structure. By analysing the ${\rm H\beta}$ lines, we find it unlikely that the C1 complex is directly related to the jet base or the accretion disk. This suggests that the Black Hole is located upstream of this complex. The projected distance between the C1 region and the jet apex is roughly $\sim$600--2400\,$R_g$, adopting the Black Hole mass of 3.2 $\times 10^8 M_\odot$. 
However, we cannot rule out a possible foreground absorber influencing our results. 

We find tentative evidence for correlations between the position angle and brightness temperatures of the C1a and C1b region and tentative evidence for a change in the viewing angle between the C1 and C3 regions of 3C\,84. If these results are confirmed, it may suggest a spatially bent jet and/or rotation of moving jet components, and indirect evidence for $\gamma$-ray emission occurring within the C1 region. A more dense millimeter-VLBI monitoring will be required to further clarify these issues.

\section*{Acknowledgements}

This research was supported by Basic Science Research Program through the National Research Foundation of Korea(NRF) funded by the Ministry of Education(NRF-2021R1A6A3A01086420).
J. A. Hodgson was supported by Korea Research Fellowship Program through the National Research Foundation of Korea(NRF) funded by the Ministry of Science and ICT(2018H1D3A1A02032824) and the research grant (2021R1C1C1009973). S.-S. Lee was supported by the National Research Foundation of Korea (NRF) grant funded by the Korea government (MIST) (2020R1A2C2009003). J.-Y. Kim is supported for this research by the International Max-Planck
Research School (IMPRS) for Astronomy and Astrophysics at the University of Bonn and Cologne. 
This research was supported by an appointment to the NASA Postdoctoral Program
at the Goddard Space Flight Center, administered by Universities Space Research Association 
through a contract with NASA. 
Rocco Lico acknowledges the support of the Spanish Ministerio de Economía y Competitividad (grant PID2019-108995GB-C21), the Consejería de Economía, Conocimiento, Empresas y Universidad of the Junta de Andalucía (grant P18-FR-1769), the Consejo Superior de Investigaciones Científicas (grant 2019AEP112). 
G. F. Paraschos is supported for this research by the International Max-Planck Research School (IMPRS) for Astronomy and Astrophysics at the University of Bonn and Cologne. 
S.T. and M.K. acknowledge support via NRF grant 2019R1F1A1059721.
This research has made use of data obtained with the Global
Millimeter VLBI Array (GMVA), which consists of telescopes operated by the
MPIfR, IRAM, Onsala, Metsahovi, Yebes, the Korean VLBI Network, the Green
Bank Observatory and the Very Long Baseline Array (VLBA). The VLBA is an
instrument of the NRAO, which is a facility of the National Science Foundation
operated under cooperative agreement by Associated Universities, Inc. The data were correlated at the correlator of the Max Planck Institute for Radioastronomy (MPIfR) in Bonn, Germany. This work made use of the Swinburne
University of Technology software correlator, developed as part of the Australian
Major National Research Facilities Programme and operated under licence.

\section*{Data Availability}

The data underlying this article will be shared on reasonable request to the corresponding author.



\bibliographystyle{mnras}
\bibliography{Bibliography} 

\begin{thebibliography}{}
\makeatletter
\relax
\def\mn@urlcharsother{\let\do\@makeother \do\$\do\&\do\#\do\^\do\_\do\%\do\~}
\def\mn@doi{\begingroup\mn@urlcharsother \@ifnextchar [ {\mn@doi@}
  {\mn@doi@[]}}
\def\mn@doi@[#1]#2{\def\@tempa{#1}\ifx\@tempa\@empty \href
  {http://dx.doi.org/#2} {doi:#2}\else \href {http://dx.doi.org/#2} {#1}\fi
  \endgroup}
\def\mn@eprint#1#2{\mn@eprint@#1:#2::\@nil}
\def\mn@eprint@arXiv#1{\href {http://arxiv.org/abs/#1} {{\tt arXiv:#1}}}
\def\mn@eprint@dblp#1{\href {http://dblp.uni-trier.de/rec/bibtex/#1.xml}
  {dblp:#1}}
\def\mn@eprint@#1:#2:#3:#4\@nil{\def\@tempa {#1}\def\@tempb {#2}\def\@tempc
  {#3}\ifx \@tempc \@empty \let \@tempc \@tempb \let \@tempb \@tempa \fi \ifx
  \@tempb \@empty \def\@tempb {arXiv}\fi \@ifundefined
  {mn@eprint@\@tempb}{\@tempb:\@tempc}{\expandafter \expandafter \csname
  mn@eprint@\@tempb\endcsname \expandafter{\@tempc}}}

\bibitem[\protect\citeauthoryear{{Abdo} et~al.,}{{Abdo}
  et~al.}{2009}]{abdo2009}
{Abdo} A.~A.,  et~al., 2009, \mn@doi [\apj] {10.1088/0004-637X/699/1/31}, \href
  {https://ui.adsabs.harvard.edu/abs/2009ApJ...699...31A} {699, 31}

\bibitem[\protect\citeauthoryear{{Baczko} et~al.,}{{Baczko}
  et~al.}{2016}]{baczko2016}
{Baczko} A.~K.,  et~al., 2016, \mn@doi [\aap] {10.1051/0004-6361/201527951},
  \href {https://ui.adsabs.harvard.edu/abs/2016A&A...593A..47B} {593, A47}

\bibitem[\protect\citeauthoryear{{Bentz} et~al.,}{{Bentz}
  et~al.}{2013}]{bentz2013}
{Bentz} M.~C.,  et~al., 2013, \mn@doi [\apj] {10.1088/0004-637X/767/2/149},
  \href {https://ui.adsabs.harvard.edu/abs/2013ApJ...767..149B} {767, 149}

\bibitem[\protect\citeauthoryear{{Blandford} \& {Payne}}{{Blandford} \&
  {Payne}}{1982}]{BP}
{Blandford} R.~D.,  {Payne} D.~G.,  1982, \mn@doi [\mnras]
  {10.1093/mnras/199.4.883}, \href
  {https://ui.adsabs.harvard.edu/abs/1982MNRAS.199..883B} {199, 883}

\bibitem[\protect\citeauthoryear{{Blandford} \& {Znajek}}{{Blandford} \&
  {Znajek}}{1977}]{BZ}
{Blandford} R.~D.,  {Znajek} R.~L.,  1977, \mn@doi [\mnras]
  {10.1093/mnras/179.3.433}, \href
  {https://ui.adsabs.harvard.edu/abs/1977MNRAS.179..433B} {179, 433}

\bibitem[\protect\citeauthoryear{{Blandford}, {Meier}  \&
  {Readhead}}{{Blandford} et~al.}{2019}]{blandford2019}
{Blandford} R.,  {Meier} D.,   {Readhead} A.,  2019, \mn@doi [\araa]
  {10.1146/annurev-astro-081817-051948}, \href
  {https://ui.adsabs.harvard.edu/abs/2019ARA&A..57..467B} {57, 467}

\bibitem[\protect\citeauthoryear{{Boccardi}, {Krichbaum}, {Bach}, {Bremer}  \&
  {Zensus}}{{Boccardi} et~al.}{2016}]{boccardi2016}
{Boccardi} B.,  {Krichbaum} T.~P.,  {Bach} U.,  {Bremer} M.,   {Zensus} J.~A.,
  2016, \mn@doi [\aap] {10.1051/0004-6361/201628412}, \href
  {https://ui.adsabs.harvard.edu/abs/2016A&A...588L...9B} {588, L9}

\bibitem[\protect\citeauthoryear{{Boccardi}, {Krichbaum}, {Ros}  \&
  {Zensus}}{{Boccardi} et~al.}{2017}]{boccardi2017}
{Boccardi} B.,  {Krichbaum} T.~P.,  {Ros} E.,   {Zensus} J.~A.,  2017, \mn@doi
  [\aapr] {10.1007/s00159-017-0105-6}, \href
  {https://ui.adsabs.harvard.edu/abs/2017A&ARv..25....4B} {25, 4}

\bibitem[\protect\citeauthoryear{{Deller} et~al.,}{{Deller}
  et~al.}{2011}]{deller2011}
{Deller} A.~T.,  et~al., 2011, \mn@doi [\pasp] {10.1086/658907}, \href
  {https://ui.adsabs.harvard.edu/abs/2011PASP..123..275D} {123, 275}

\bibitem[\protect\citeauthoryear{{Dhawan}, {Kellermann}  \& {Romney}}{{Dhawan}
  et~al.}{1998}]{dhawan1998}
{Dhawan} V.,  {Kellermann} K.~I.,   {Romney} J.~D.,  1998, \mn@doi [\apjl]
  {10.1086/311313}, \href
  {https://ui.adsabs.harvard.edu/abs/1998ApJ...498L.111D} {498, L111}

\bibitem[\protect\citeauthoryear{{Fomalont}}{{Fomalont}}{1999}]{fomalont1999}
{Fomalont} E.~B.,  1999, in {Taylor} G.~B.,  {Carilli} C.~L.,   {Perley} R.~A.,
   eds,  Astronomical Society of the Pacific Conference Series Vol. 180,
  Synthesis Imaging in Radio Astronomy II. p.~301

\bibitem[\protect\citeauthoryear{{Fujita} \& {Nagai}}{{Fujita} \&
  {Nagai}}{2017}]{fujita2017}
{Fujita} Y.,  {Nagai} H.,  2017, \mn@doi [\mnras] {10.1093/mnrasl/slw217},
  \href {https://ui.adsabs.harvard.edu/abs/2017MNRAS.465L..94F} {465, L94}

\bibitem[\protect\citeauthoryear{{Ghisellini}, {Tavecchio}  \&
  {Chiaberge}}{{Ghisellini} et~al.}{2005}]{ghisellini2005}
{Ghisellini} G.,  {Tavecchio} F.,   {Chiaberge} M.,  2005, \mn@doi [\aap]
  {10.1051/0004-6361:20041404}, \href
  {https://ui.adsabs.harvard.edu/abs/2005A&A...432..401G} {432, 401}

\bibitem[\protect\citeauthoryear{{Giovannini} et~al.,}{{Giovannini}
  et~al.}{2018}]{giovannini2018}
{Giovannini} G.,  et~al., 2018, \mn@doi [Nature Astronomy]
  {10.1038/s41550-018-0431-2}, \href
  {https://ui.adsabs.harvard.edu/abs/2018NatAs...2..472G} {2, 472}

\bibitem[\protect\citeauthoryear{{Greene} \& {Ho}}{{Greene} \&
  {Ho}}{2005}]{greene2005}
{Greene} J.~E.,  {Ho} L.~C.,  2005, \mn@doi [\apj] {10.1086/431897}, \href
  {https://ui.adsabs.harvard.edu/abs/2005ApJ...630..122G} {630, 122}

\bibitem[\protect\citeauthoryear{{Hiura} et~al.,}{{Hiura}
  et~al.}{2018}]{hiura2018}
{Hiura} K.,  et~al., 2018, \mn@doi [\pasj] {10.1093/pasj/psy078}, \href
  {https://ui.adsabs.harvard.edu/abs/2018PASJ...70...83H} {70, 83}

\bibitem[\protect\citeauthoryear{{Hodgson} et~al.,}{{Hodgson}
  et~al.}{2017}]{hodgson2017}
{Hodgson} J.~A.,  et~al., 2017, \mn@doi [\aap] {10.1051/0004-6361/201526727},
  \href {https://ui.adsabs.harvard.edu/abs/2017A&A...597A..80H} {597, A80}

\bibitem[\protect\citeauthoryear{{Hodgson}, {Rani}  \& {Oh}}{{Hodgson}
  et~al.}{2018a}]{hodgson2018a}
{Hodgson} J.~A.,  {Rani} B.,   {Oh} J.,  2018a, arXiv e-prints, \href
  {https://ui.adsabs.harvard.edu/abs/2018arXiv180202766H} {p. arXiv:1802.02766}

\bibitem[\protect\citeauthoryear{{Hodgson} et~al.,}{{Hodgson}
  et~al.}{2018b}]{hodgson2018b}
{Hodgson} J.~A.,  et~al., 2018b, \mn@doi [\mnras] {10.1093/mnras/stx3041},
  \href {https://ui.adsabs.harvard.edu/abs/2018MNRAS.475..368H} {475, 368}

\bibitem[\protect\citeauthoryear{{Hodgson} et~al.,}{{Hodgson}
  et~al.}{2021}]{hodgson2021}
{Hodgson} J.~A.,  et~al., 2021, \mn@doi [\apj] {10.3847/1538-4357/abf6dd},
  \href {https://ui.adsabs.harvard.edu/abs/2021ApJ...914...43H} {914, 43}

\bibitem[\protect\citeauthoryear{{H{\"o}gbom}}{{H{\"o}gbom}}{1974}]{hogbom1974}
{H{\"o}gbom} J.~A.,  1974, \aaps, \href
  {https://ui.adsabs.harvard.edu/abs/1974A&AS...15..417H} {15, 417}

\bibitem[\protect\citeauthoryear{{Hovatta}, {Valtaoja}, {Tornikoski}  \&
  {L{\"a}hteenm{\"a}ki}}{{Hovatta} et~al.}{2009}]{hovatta09}
{Hovatta} T.,  {Valtaoja} E.,  {Tornikoski} M.,   {L{\"a}hteenm{\"a}ki} A.,
  2009, \mn@doi [\aap] {10.1051/0004-6361:200811150}, \href
  {https://ui.adsabs.harvard.edu/abs/2009A&A...494..527H} {494, 527}

\bibitem[\protect\citeauthoryear{{Jorstad} et~al.,}{{Jorstad}
  et~al.}{2005}]{jorstad2005}
{Jorstad} S.~G.,  et~al., 2005, \mn@doi [\aj] {10.1086/444593}, \href
  {https://ui.adsabs.harvard.edu/abs/2005AJ....130.1418J} {130, 1418}

\bibitem[\protect\citeauthoryear{{Jorstad} et~al.,}{{Jorstad}
  et~al.}{2017}]{jor17}
{Jorstad} S.~G.,  et~al., 2017, \mn@doi [\apj] {10.3847/1538-4357/aa8407},
  \href {https://ui.adsabs.harvard.edu/abs/2017ApJ...846...98J} {846, 98}

\bibitem[\protect\citeauthoryear{{Kim} et~al.,}{{Kim} et~al.}{2019}]{kim2019}
{Kim} J.~Y.,  et~al., 2019, \mn@doi [\aap] {10.1051/0004-6361/201832920}, \href
  {https://ui.adsabs.harvard.edu/abs/2019A&A...622A.196K} {622, A196}

\bibitem[\protect\citeauthoryear{{Krichbaum} et~al.,}{{Krichbaum}
  et~al.}{1992}]{krichbaum1992}
{Krichbaum} T.~P.,  et~al., 1992, \aap, \href
  {https://ui.adsabs.harvard.edu/abs/1992A&A...260...33K} {260, 33}

\bibitem[\protect\citeauthoryear{{Lee}, {Lobanov}, {Krichbaum}, {Witzel},
  {Zensus}, {Bremer}, {Greve}  \& {Grewing}}{{Lee} et~al.}{2008}]{lee2008}
{Lee} S.-S.,  {Lobanov} A.~P.,  {Krichbaum} T.~P.,  {Witzel} A.,  {Zensus} A.,
  {Bremer} M.,  {Greve} A.,   {Grewing} M.,  2008, \mn@doi [\aj]
  {10.1088/0004-6256/136/1/159}, \href
  {https://ui.adsabs.harvard.edu/abs/2008AJ....136..159L} {136, 159}

\bibitem[\protect\citeauthoryear{{Lico}, {Giroletti}, {Orienti}  \&
  {D'Ammando}}{{Lico} et~al.}{2016}]{lico2016}
{Lico} R.,  {Giroletti} M.,  {Orienti} M.,   {D'Ammando} F.,  2016, \mn@doi
  [\aap] {10.1051/0004-6361/201628775}, \href
  {https://ui.adsabs.harvard.edu/abs/2016A&A...594A..60L} {594, A60}

\bibitem[\protect\citeauthoryear{{Liodakis} et~al.,}{{Liodakis}
  et~al.}{2017}]{liodakis17}
{Liodakis} I.,  et~al., 2017, \mn@doi [\mnras] {10.1093/mnras/stx002}, \href
  {https://ui.adsabs.harvard.edu/abs/2017MNRAS.466.4625L} {466, 4625}

\bibitem[\protect\citeauthoryear{{Lister} et~al.,}{{Lister}
  et~al.}{2009}]{lister2009}
{Lister} M.~L.,  et~al., 2009, \mn@doi [\aj] {10.1088/0004-6256/138/6/1874},
  \href {https://ui.adsabs.harvard.edu/abs/2009AJ....138.1874L} {138, 1874}

\bibitem[\protect\citeauthoryear{{Nagai} et~al.,}{{Nagai}
  et~al.}{2014}]{nagai2014}
{Nagai} H.,  et~al., 2014, \mn@doi [\apj] {10.1088/0004-637X/785/1/53}, \href
  {https://ui.adsabs.harvard.edu/abs/2014ApJ...785...53N} {785, 53}

\bibitem[\protect\citeauthoryear{{Onori} et~al.,}{{Onori}
  et~al.}{2017}]{onori2017}
{Onori} F.,  et~al., 2017, \mn@doi [\mnras] {10.1093/mnrasl/slx032}, \href
  {https://ui.adsabs.harvard.edu/abs/2017MNRAS.468L..97O} {468, L97}

\bibitem[\protect\citeauthoryear{{Paraschos}, {Kim}, {Krichbaum}  \&
  {Zensus}}{{Paraschos} et~al.}{2021}]{paraschos2021}
{Paraschos} G.~F.,  {Kim} J.~Y.,  {Krichbaum} T.~P.,   {Zensus} J.~A.,  2021,
  \mn@doi [\aap] {10.1051/0004-6361/202140776}, \href
  {https://ui.adsabs.harvard.edu/abs/2021A&A...650L..18P} {650, L18}

\bibitem[\protect\citeauthoryear{{Park} \& {Trippe}}{{Park} \&
  {Trippe}}{2017}]{park2017}
{Park} J.,  {Trippe} S.,  2017, \mn@doi [\apj] {10.3847/1538-4357/834/2/157},
  \href {https://ui.adsabs.harvard.edu/abs/2017ApJ...834..157P} {834, 157}

\bibitem[\protect\citeauthoryear{{Planck Collaboration} et~al.,}{{Planck
  Collaboration} et~al.}{2016}]{planck}
{Planck Collaboration} et~al., 2016, \mn@doi [\aap]
  {10.1051/0004-6361/201525830}, \href
  {https://ui.adsabs.harvard.edu/abs/2016A&A...594A..13P} {594, A13}

\bibitem[\protect\citeauthoryear{{Punsly}, {Marziani}, {Bennert}, {Nagai}  \&
  {Gurwell}}{{Punsly} et~al.}{2018}]{punsly2018}
{Punsly} B.,  {Marziani} P.,  {Bennert} V.~N.,  {Nagai} H.,   {Gurwell} M.~A.,
  2018, \mn@doi [\apj] {10.3847/1538-4357/aaec75}, \href
  {https://ui.adsabs.harvard.edu/abs/2018ApJ...869..143P} {869, 143}

\bibitem[\protect\citeauthoryear{{Punsly}, {Nagai}, {Savolainen}  \&
  {Orienti}}{{Punsly} et~al.}{2021}]{punsly2021}
{Punsly} B.,  {Nagai} H.,  {Savolainen} T.,   {Orienti} M.,  2021, \mn@doi
  [\apj] {10.3847/1538-4357/abe69f}, \href
  {https://ui.adsabs.harvard.edu/abs/2021ApJ...911...19P} {911, 19}

\bibitem[\protect\citeauthoryear{{Readhead}}{{Readhead}}{1994}]{readhead1994}
{Readhead} A. C.~S.,  1994, \mn@doi [\apj] {10.1086/174038}, \href
  {https://ui.adsabs.harvard.edu/abs/1994ApJ...426...51R} {426, 51}

\bibitem[\protect\citeauthoryear{{Scharw{\"a}chter}, {McGregor}, {Dopita}  \&
  {Beck}}{{Scharw{\"a}chter} et~al.}{2013}]{scharwachter2013}
{Scharw{\"a}chter} J.,  {McGregor} P.~J.,  {Dopita} M.~A.,   {Beck} T.~L.,
  2013, \mn@doi [\mnras] {10.1093/mnras/sts502}, \href
  {https://ui.adsabs.harvard.edu/abs/2013MNRAS.429.2315S} {429, 2315}

\bibitem[\protect\citeauthoryear{{Shepherd}, {Pearson}  \& {Taylor}}{{Shepherd}
  et~al.}{1994}]{shepherd1994}
{Shepherd} M.~C.,  {Pearson} T.~J.,   {Taylor} G.~B.,  1994, in Bulletin of the
  American Astronomical Society. pp 987--989

\bibitem[\protect\citeauthoryear{{Singal}}{{Singal}}{2009}]{singal2009}
{Singal} A.~K.,  2009, \mn@doi [\apjl] {10.1088/0004-637X/703/2/L109}, \href
  {https://ui.adsabs.harvard.edu/abs/2009ApJ...703L.109S} {703, L109}

\bibitem[\protect\citeauthoryear{Stetson}{Stetson}{1997}]{stetson1997}
Stetson P.~B.,  1997, \mn@doi [PASP] {10.1086/131977}, \href
  {https://ui.adsabs.harvard.edu/abs/1987PASP...99..191S/abstract} {99, 191}

\bibitem[\protect\citeauthoryear{{Strauss}, {Huchra}, {Davis}, {Yahil},
  {Fisher}  \& {Tonry}}{{Strauss} et~al.}{1992}]{strauss1992}
{Strauss} M.~A.,  {Huchra} J.~P.,  {Davis} M.,  {Yahil} A.,  {Fisher} K.~B.,
  {Tonry} J.,  1992, \mn@doi [\apjs] {10.1086/191730}, \href
  {https://ui.adsabs.harvard.edu/abs/1992ApJS...83...29S} {83, 29}

\bibitem[\protect\citeauthoryear{{Suzuki} et~al.,}{{Suzuki}
  et~al.}{2012}]{suzuki2012}
{Suzuki} K.,  et~al., 2012, \mn@doi [\apj] {10.1088/0004-637X/746/2/140}, \href
  {https://ui.adsabs.harvard.edu/abs/2012ApJ...746..140S} {746, 140}

\bibitem[\protect\citeauthoryear{{Wajima}, {Kino}  \& {Kawakatu}}{{Wajima}
  et~al.}{2020}]{wajima2020}
{Wajima} K.,  {Kino} M.,   {Kawakatu} N.,  2020, \mn@doi [\apj]
  {10.3847/1538-4357/ab88a0}, \href
  {https://ui.adsabs.harvard.edu/abs/2020ApJ...895...35W} {895, 35}

\bibitem[\protect\citeauthoryear{{Walker}, {Dhawan}, {Romney}, {Kellermann}  \&
  {Vermeulen}}{{Walker} et~al.}{2000}]{walker2000}
{Walker} R.~C.,  {Dhawan} V.,  {Romney} J.~D.,  {Kellermann} K.~I.,
  {Vermeulen} R.~C.,  2000, \mn@doi [\apj] {10.1086/308372}, \href
  {https://ui.adsabs.harvard.edu/abs/2000ApJ...530..233W} {530, 233}

\makeatother
\end{thebibliography}








\bsp	
\label{lastpage}
\end{document}